\documentstyle[epsfig,wrapfig]{article}
\newcommand{\be}{\begin{equation}}
\newcommand{\ee}{\end{equation}}
\newcommand{\ba}{\begin{eqnarray}}
\newcommand{\ea}{\end{eqnarray}}
\newcommand{\la}{\langle}
\newcommand{\ra}{\rangle}
\newcommand{\di}{ {\rm d} }
\newcommand{\SIDIS}{{\mbox{\tiny SIDIS}}}
\newcommand{\DY   }{{\mbox{\tiny DY}}}

\begin{document}
\title{\Large\bf Sivers effect in semi-inclusive DIS and in the Drell-Yan process}
\author{
  A.~V.~Efremov$^a$,
  K.~Goeke$^b$,
  S.~Menzel$^b$,
  A.~Metz$^b$,
  P.~Schweitzer$^b$ \\
  \footnotesize\it $^a$
  Joint Institute for Nuclear Research, Dubna, 141980 Russia\\
  \footnotesize\it $^b$
  Institut f\"ur Theoretische Physik II, Ruhr-Universit\"at Bochum,
  Germany}
\date{}
\maketitle
\begin{abstract}
\noindent
The Sivers parton distribution function has been predicted to obey a particular
``universality relation'', namely to have opposite sign in semi-inclusive
deeply inelastic scattering (SIDIS) and the Drell-Yan process.
We discuss how, on the basis of present HERMES data, this remarkable
prediction of the QCD factorization approach to the description of single
spin asymmetries related to the Sivers effect could be checked
experimentally in future experiments at PAX and COMPASS.
\end{abstract}

\section{Introduction}
\label{Sec-1:introduction}

It was understood early \cite{Kane:nd} that single spin asymmetries (SSA)
in hard processes, such as those observed in $p^\uparrow p\to \pi X$
\cite{Adams:1991rw,Adams:2003fx} or in SIDIS
\cite{Airapetian:1999tv,Avakian:2003pk,Avetisyan:2004uz,HERMES-new,Airapetian:2004tw,Bradamante:2004qh},
cannot be explained by means of leading twist collinear QCD factorization.
One of the non-perturbative effects which could account for such SSA 
considers a non-trivial correlation between (the transverse component of) 
the nucleon spin ${\bf S}_{\rm T}$ and intrinsic transverse parton momenta 
${\bf p}_{\rm T}$ in the nucleon \cite{Sivers:1989cc}, and is quantified 
in terms of the so-called Sivers function $f_{1T}^\perp(x,{\bf p}_T^2)$ 
\cite{Boer:1997nt}.
The effect is referred to as ``(naively) T-odd'', since it is proportional,
e.g., in the infinite momentum frame where the nucleon momentum
${\bf P}_N\to\infty$, to the T-odd structure
$({\bf S}_{\rm T}\times{\bf p}_{\rm T}){\bf P}_N$.
The Sivers effect was shown to be able to explain the SSA in
$p^\uparrow p\to \pi X$ \cite{Anselmino:1994tv}, though also
other mechanisms exist which could contribute in this reaction
\cite{Efremov:eb,Collins:1992kk,Kanazawa:2000hz}.

The precise definition of $f_{1T}^\perp(x,{\bf p}_T^2)$ in QCD was worked
out only recently \cite{Brodsky:2002cx,Collins:2002kn,Belitsky:2002sm}.
A particularly interesting feature of the Sivers function concerns its
universality property. This property ensures for usual parton distributions
that one deals with, e.g., the same unpolarized parton distribution $f_1(x)$
in SIDIS and in the Drell-Yan process (DY):
$f_1(x)_\SIDIS = f_1(x)_\DY$.
In the case of the Sivers function (and other T-odd distributions) the
universality property takes, however, a different form. On the basis of
time-reversal arguments it was predicted \cite{Collins:2002kn} that
$f_{1T}^\perp$ in SIDIS and DY have opposite sign,
\be\label{Eq:01}
     f_{1T}^\perp(x,{\bf p}_T^2)_\SIDIS =
    -f_{1T}^\perp(x,{\bf p}_T^2)_\DY \;.
\ee

The experimental check of Eq.~(\ref{Eq:01}) would provide a thorough test
of our understanding of the Sivers effect within QCD and, hence, our
understanding of SSA.
It would crucially test the factorization approach to the description of
processes sensitive to transverse parton
momenta \cite{Collins:1981uk,Ji:2004wu,Collins:2004nx}.

In this work we shall discuss how the relation (\ref{Eq:01}) could be checked
experimentally in the {\bf P}olarized {\bf A}ntiproton e{\bf X}periment (PAX)
planned at GSI \cite{PAX,Rathmann:2004pm}. A primary goal of this experiment
will be to provide a polarized antiproton beam and to measure the transversity
distribution $h_1^a(x)$, c.f.\ \cite{PAX-estimates}.
However, PAX will also be well suited to access the Sivers function via SSA
in $\bar{p}p^\uparrow\to\mu^+\mu^-X$ or $\bar{p}^\uparrow p\to\mu^+\mu^-X$
\cite{PAX,Rathmann:2004pm}. In the COMPASS experiment at CERN \cite{proposal},
making use of a $\pi^-$ beam, one would also be able to study the Sivers
function via SSA in $\pi^-p^\uparrow\to\mu^+\mu^-X$.

In order to estimate the magnitude of the Sivers effect in those experiments 
we will roughly parameterize $f_{1T}^\perp(x,{\bf p}_T^2)_\SIDIS$ from 
the (preliminary) HERMES data \cite{HERMES-new} using as a guideline 
relations derived from the QCD limit of a large number of colours $N_c$ 
\cite{Pobylitsa:2003ty}. Such large-$N_c$ relations are observed to hold 
in nature within their expected accuracy \cite{Efremov:2000ar} and, as a 
byproduct of our study, we shall observe that this is also the case here.
On the basis of the obtained parameterization we estimate SSA 
in the Drell-Yan process for the PAX and COMPASS experiments. 
We also comment briefly on parameterizations of $f_{1T}^\perp$ reported 
previously in the literature and on model calculations.

\section{The Sivers function}
\label{Sec-2:Sivers-function}

A definition of the unintegrated unpolarized distribution function
$f_1(x,{\bf p}_T^2)$ and the Sivers function $f_{1T}^\perp(x,{\bf p}_T^2)$
can be given in terms of the light-cone correlator
\ba
    \Phi^q(x,{\bf p}_T) &\equiv&
    \int\!\frac{\di\xi^-\di^2\xi_T\!\!}{2(2\pi)^3}\,e^{ip\cdot\xi}
    \la P,S_T|\bar{\psi}_q(0) \gamma_\mu n_-^\mu \;
    {\cal W}[0,\xi;{\rm process}]\;
    \psi_q(\xi)|P,S_T\ra\biggl|_{\xi^+=0} \nonumber\\
    &=& f_1^q(x,{\bf p}_T^2) + f_{1T}^{\perp q}(x,{\bf p}_T^2)\;\frac{
    \varepsilon_{\mu\nu\rho\sigma}n_-^\mu n_+^\nu p_T^\rho S_T^\sigma}{M_N} \;,
    \label{Eq:th-01}\ea
where the dimensionless light-like vectors $n_\pm$ are defined such that
$n_+\cdot n_-=1$. (See Ref.~\cite{Collins:2003fm} for a precise definition
and the meaning of unintegrated distribution functions in QCD.)

\begin{figure}[t!]
\begin{center}
        \includegraphics[width=12cm]{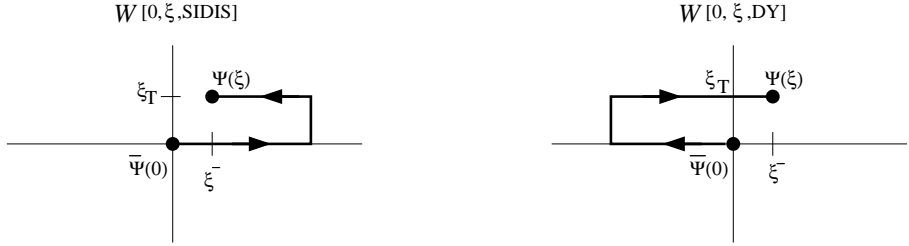}
    \caption{\footnotesize\sl
    The path of the process-dependent gauge link
    ${\cal W}[0,\xi;\mbox{process}]$ which enters the
    definition of the Sivers function in SIDIS and DY.
    \label{fig1-processes-diagramms}}
\end{center}
\end{figure}

The Wilson link ${\cal W}[0,\xi;{\rm process}]$
is defined in Fig.~\ref{fig1-processes-diagramms}, c.f.\
Refs.~\cite{Collins:2002kn,Belitsky:2002sm}.
For observables integrated over ${\bf p}_T$ the process dependence of the
gauge link usually cancels out. However, the situation is different for 
$f_{1T}^\perp$. If one neglected the gauge link, under time-reversal the 
Sivers function would transform into its negative, i.e.\  it would vanish 
\cite{Collins:1992kk}. However, initial or final state interactions
\cite{Brodsky:2002cx,Afanasev:2003ze}, needed to obtain non-zero SSA 
\cite{Gasiorowicz}, generate a Wilson link for the Sivers function in 
any gauge \cite{Collins:2002kn,Belitsky:2002sm}.
Under time reversal the gauge link of SIDIS is transformed into the
gauge link of DY, and vice versa \cite{Collins:2002kn}.
(The gauge link structure in the more involved hadronic process
$p^\uparrow p\to\pi X$  was addressed in \cite{Bomhof:2004aw}.)
This yields the peculiar universality relation in Eq.~(\ref{Eq:01}).

Only little is known about the non-perturbative properties of the Sivers
function. In Ref.~\cite{Bacchetta:1999kz} bounds were derived from the
positivity of the spin density matrix, which constrain $f_{1T}^\perp$
in terms of other transverse momentum dependent distributions
including also so far experimentally unknown functions. Eliminating the
unknown distribution, at the price of relaxing the bound, one obtains
\be\label{Eq:th-02}
    \frac{|{\bf p}_T|}{M_N}\;|f_{1T}^{\perp q}(x,{\bf p}_T^2)|
    \le f_1^q(x,{\bf p}_T^2) \;.
\ee

The average parton transverse momentum defined as
$\la{\bf p}_T^q\ra=$
$\int\!\di x\int\!\di^2{\bf p}_T \;{\bf p}_T\Phi^q(x,{\bf p}_T)$
was shown ~\cite{Burkardt:2003yg,Burkardt:2004ur}
(with analogously defined gluon transverse momentum) to obey the relation
\be\label{Eq:th-03}
    \sum_{a=g,u,d,\,\dots} \la{\bf p}_T^a\ra = 0 \;.
\ee
Eq.~(\ref{Eq:th-03}) is more than trivial momentum conservation in the
plane transverse to the hard momentum flow, since it connects transverse
momenta due to final state interactions which the scattered quark experiences
in incoherently summed scattering events
\cite{Burkardt:2003yg,Burkardt:2004ur}.
Inserting the definition of $\Phi^q(x,{\bf p}_T)$ into Eq.~(\ref{Eq:th-03})
and using the fact that (analogously for gluons)
%
\be\label{Eq:th-04}
    \int\!\di x\int\!\di^2{\bf p}_T \;\frac{p_T^k p_T^l}{M_N^2}
    \,f_{1T}^{\perp q}(x,{\bf p}_T^2)
    = \delta^{kl}\int\!\di x \;f_{1T}^{\perp(1)q}(x) \;,
    \;\;\;\mbox{with}\;\;\;
    f_{1T}^{\perp(1)q}(x)\equiv \int\!\di^2{\bf p}_T \;
    \frac{{\bf p}_T^2}{2M_N^2}\,f_{1T}^{\perp q}(x,{\bf p}_T^2) \;,
\ee
one obtains
\be\label{Eq:th-05}
    \sum_{a=g,u,d,\,\dots}
    \int\!\di x \;f_{1T}^{\perp(1)a}(x) = 0\;.
\ee
The sum rule (\ref{Eq:th-05}) may prove a useful constraint for
parameterizations of the Sivers function, in particular, because it
is $f_{1T}^{\perp(1)}(x)$ which enters in a model independent way cross
sections properly weighted with transverse momenta \cite{Boer:1997nt}.

Another property of the Sivers function, which will be used later on,
is the relation derived in the limit of a large number of colours $N_c$
in QCD \cite{Pobylitsa:2003ty}, namely
\be\label{Eq:th-06}
      f_{1T}^{\perp u}(x,{\bf p}_T^2) =
    - f_{1T}^{\perp d}(x,{\bf p}_T^2) \;\;\;
    \mbox{modulo $1/N_c$ corrections.}
\ee
It should be noted that in the large-$N_c$ limit $xN_c={\cal O}(1)$, such
that Eq.~(\ref{Eq:th-06}) can be expected to be satisfied in the valence
region of not too small and not too large $x$ to within an accuracy of
${\cal O}(1/N_c)$ \cite{Efremov:2000ar}.

Neglecting strange and heavier quarks, which is a reasonable
assumption in the case of the nucleon, one obtains from
Eqs.~(\ref{Eq:th-05},~\ref{Eq:th-06}) that the Sivers gluon
distribution is suppressed in the large-$N_c$ limit with respect
to the quark distribution functions. More precisely, it is of the
same order of magnitude as the flavour singlet combination\footnote{
    What matters in the large $N_c$-counting is the spin-flavour
    symmetry of the involved operator. In this respect the operator
    entering the flavour singlet and the gluon Sivers function
    have the same behaviour and thus the same large-$N_c$ counting.
    We thank Pavel Pobylitsa for discussions on this point.}, i.e., 
$(f_{1T}^{\perp u}+f_{1T}^{\perp d})\sim f_{1T}^{\perp g}\sim{\cal O}(N_c^2)$.
Thus, in the large-$N_c$ limit the gluon Sivers effect can be expected to be
suppressed with respect to the non-singlet quark Sivers effect at not too
small $x$, which is an interesting constraint for phenomenological studies.
In order to obtain a feeling to which extent such large-$N_c$ relations 
may be expected to hold, it is interesting to mention that the helicity 
distribution function exhibits a similar behaviour in the large-$N_c$
limit, namely $|(g_1^u-g_1^d)(x)|\sim{\cal O}(N_c^2)$ is larger than
$|(g_1^u+g_1^d)(x)| \sim |g_1^g(x)|\sim{\cal O}(N_c)$.
This is -- for quarks -- roughly consistent with phenomenology, and
predicts a suppression of the (presently poorly known) helicity gluon 
distribution with respect to the unpolarized gluon distribution
function $g_1^g(x)/f_1^g(x) \sim 1/N_c$ at moderate values of $x$
\cite{Efremov:2000ar}.

The Sivers function was studied in several models, in which the gauge 
link was modeled explicitly by considering the perturbative effect 
of one-gluon exchange. Calculations based on the spectator model 
\cite{Brodsky:2002cx,Bacchetta:2003rz,Lu:2004au} and the bag model 
\cite{Yuan:2003wk} yield a sizeable $f_{1T}^{\perp u}$ but a negligible 
$f_{1T}^{\perp d}$ and, thus, for the chosen parameter sets, do not 
respect the large $N_c$ counting rule in Eq.~(\ref{Eq:th-06}).

In a large class of chiral models, which are based on Goldstone boson 
and (effective) quark degrees of freedom, T-odd distributions are zero 
\cite{Pobylitsa:2002fr}. This can be understood by recalling that in 
such models there are no gluons, whose presence is crucial to generate 
T-odd effects via the gauge link structure.
Combining the no-go-theorem (concerning modeling of $f_{1T}^\perp$ in
chiral models) of Ref.~\cite{Pobylitsa:2002fr} with notions from the
instanton model of the QCD vacuum \cite{Diakonov:1995qy}
one is lead to the suspicion \cite{Efremov:2003tf} that T-odd distributions 
could be suppressed with respect to T-even distributions in the instanton 
vacuum model, which is supported by estimates \cite{Ostrovsky:2004pd}.
For a discussion of possible instanton effects in the Drell-Yan process
see \cite{Boer:2004mv} and references therein.

By assuming that the SSA in $p^\uparrow p\to \pi X$
\cite{Adams:1991rw} is dominated by the Sivers effect
phenomenological parameterizations of the Sivers function were obtained
\cite{Anselmino:1994tv,Anselmino:1998yz,D'Alesio:2004up} which, worthwhile
stressing, approximately respect the large-$N_c$ pattern in
Eq.~(\ref{Eq:th-06}).
The assumption that this process is dominated by the Sivers effect
seems to be reasonable in the light of recent studies
\cite{Anselmino:2004ky,Ma:2004tr}, which show that the contribution
of the Collins effect in this process is small. However, one also has to
keep in mind twist-3 effects \cite{Efremov:eb,Kanazawa:2000hz}
which might be equally important.

Let us finally mention that a relation of $f_{1T}^{\perp q}$ to 
the generalized parton distribution $E^q(x,\xi,t)$ was proposed, namely 
%
%
the leading light-cone Fock component of the Sivers function may be
represented as a convolution of the Wilson link and the same overlap 
integrals between light-cone wave functions differing by one unit 
of orbital angular momentum, which enter the description of $E^q(x,\xi,t)$
\cite{Brodsky:2002cx,Burkardt:2002ks}. This has been seen explicitly 
in a quark-diquark model calculation \cite{Burkardt:2003je} and is compatible 
with the large $N_c$-limit in the sense that $E^q$ and $f_{1T}^{\perp q}$ 
have the same large-$N_c$ behaviour \cite{Petrov:1998kf}.
 From these relations it was concluded that
\be\label{Eq:th-07}
    \int\!\di x\;f_{1T\,\SIDIS}^{\perp(1) u}(x) < 0 \;,\;\;\;
    \int\!\di x\;f_{1T\,\SIDIS}^{\perp(1) d}(x) > 0 \;.
\ee
The above connection to the generalized parton distribution could further be
exploited to draw conclusions on the large-$x$ behaviour of the Sivers 
function. 
Since $E^q(x,\xi,t)\propto(1-x)^5$ at large $x$ \cite{Burkardt:2003ck}
one may conclude that also $f_{1T}^{(1)\perp q}(x) \propto (1-x)^5$. 
%
%
One must keep in mind, however, that the above assumes dimensional 
counting behaviour of the usual quark distribution functions, which yields 
$f_1^q(x)\propto(1-x)^3$ at large $x$ \cite{Farrar:1975yb}. While for 
$f_1^u(x)$ parameterizations, e.g.\ \cite{Gluck:1998xa,Martin:2002dr}, are 
roughly compatible with this prediction, this is not the case for $f_1^d(x)$.
However, if this were true also for $E^q(x,\xi,t)$ and 
$f_{1T}^{(1)\perp q}(x)$, there need not to be a conflict with large-$N_c$ 
relations, such as Eq.~(\ref{Eq:th-06}), which apply only as long as 
$xN_c={\cal O}(1)$.

\newpage
\section{Sivers effect in SIDIS}
\label{Sec-3:Sivers-in-SIDIS}

\begin{figure}[t]
    \begin{tabular}{cc}
    \hspace{1cm}
    {\epsfxsize=2.4in\epsfbox{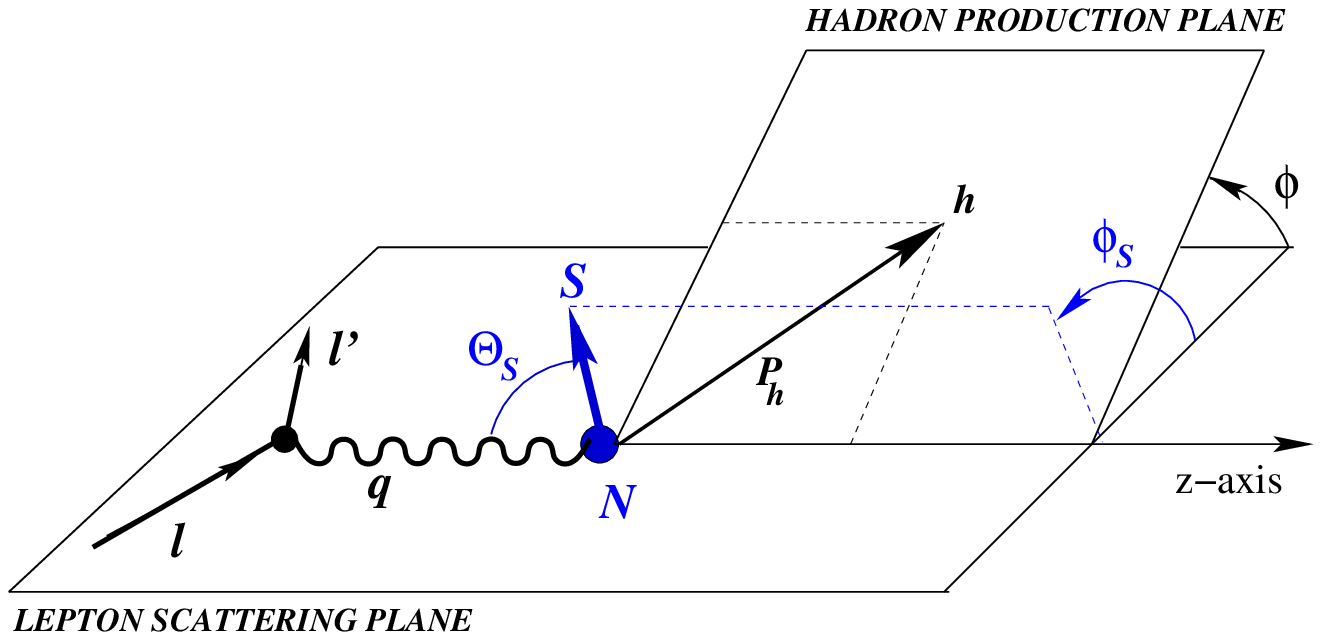}}&
    {\epsfxsize=3.0in\epsfbox{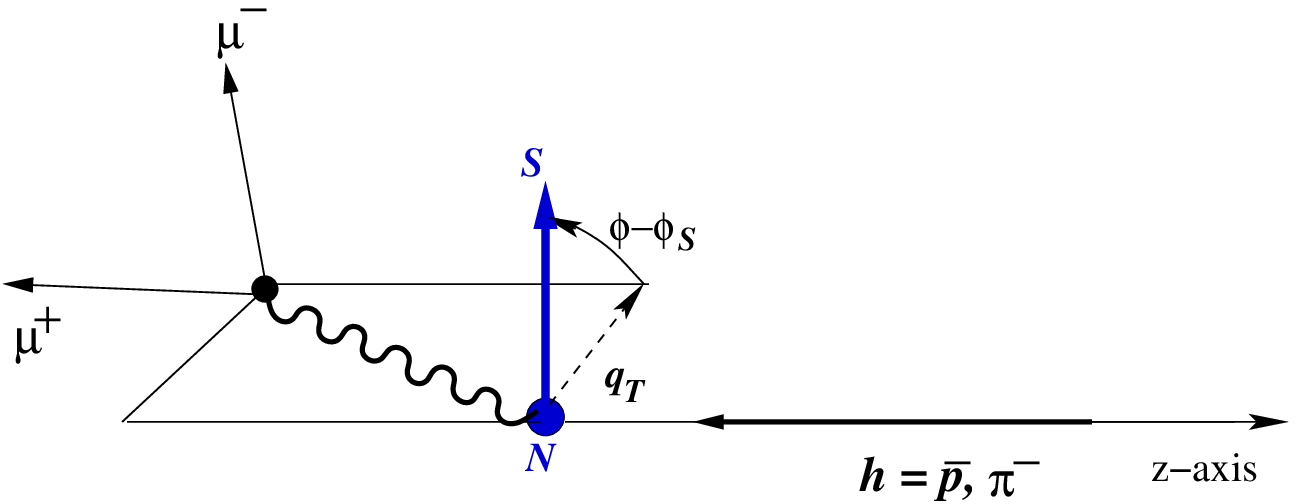}}
    \end{tabular}
        \caption{\label{fig2-processes-kinematics}\footnotesize\sl
    Kinematics of the SIDIS process $lp\rightarrow l^\prime h X$
    (left),
    and the Drell-Yan process $p^{\uparrow}h \to l^{+}l^{-} X$
    (right) in the lab frame.}
    \vspace{-0.3cm}
        \end{figure}

Consider the process $lp^\uparrow\rightarrow l'h X$ (see
Fig.~\ref{fig2-processes-kinematics}) where
``$^\uparrow$'' denotes the transverse (with respect to the beam) target
polarization. Let $P$, $l$ ($l^\prime$) and $P_h$ denote respectively the
momentum of the target proton, incoming (outgoing) lepton and the produced
hadron. The relevant kinematic variables are $s:=(P+l)^2$,
$q= l-l'$ with $Q^2=- q^2$, $x=\frac{Q^2}{2Pq}$, $y = \frac{Pq}{Pl}$,
$z=\frac{PP_h}{Pq}$.
The azimuthal SSA asymmetry is defined as
\be\label{Eq:03}
    A_{UT}^{\sin(\phi-\phi_S)\frac{P_{h\perp}}{ M_N}}(x) =
    \frac{\sum_i\sin(\phi_i-\phi_{S,i})\frac{|{\bf P}_{h\perp,i}|}{M_N}
    \{N^\uparrow(\phi_i,\phi_{S,i})-N^\downarrow(\phi_i,\phi_{S,i}+\pi)\}}
    {\frac12\sum_i
    \{N^\uparrow(\phi_i,\phi_{S,i})+N^\downarrow(\phi_i,\phi_{S,i}+\pi)\}}
    \;, \ee
where $N^{\uparrow(\downarrow)}(\phi_i,\phi_{S,i})$ are the event counts for 
the respective target polarization (corrected for depolarization effects).
The z-axis is chosen in direction of the virtual photon
(see Fig.~\ref{fig2-processes-kinematics}), in agreement with the HERMES
convention \cite{Airapetian:1999tv}.
The angles $\phi_h$ and $\phi_S$ are the azimuthal angles of the produced
hadron and the target spin. Neglecting power suppressed terms
$\propto M_N^2/Q^2$ the SSA (\ref{Eq:03}) is given by
\cite{Boer:1997nt}\footnote{Note that throughout this paper we also neglect
           contributions from soft
           gluons \cite{Collins:1981uk,Ji:2004wu,Collins:2004nx}.}
\be\label{Eq:04}
    A_{UT,\pi}^{\sin(\phi-\phi_S)\frac{P_{h\perp}}{ M_N}}(x)= -2 \;\,\frac{
    \displaystyle
    \int_{\rm cuts}\!\!\!\!\di z\,\di y\;\frac{4\pi\alpha^2 s}{Q^4}
    \biggl(1-y+\frac{y^2}{2}\biggr)
    \sum_a e_a^2\,x f_{1T\,\SIDIS}^{\perp(1)a}(x)\,z D_1^{a/\pi}(z)}{
    \displaystyle \;
    \int_{\rm cuts}\!\!\!\!\di z\,\di y\;\frac{4\pi\alpha^2 s}{Q^4}\,
    \biggl(1-y+\frac{y^2}{2}\biggr)
    \sum_a e_a^2\,x f_1^a(x)\,D_1^{a/\pi}(z)} \;.
\ee

%
%

There are two reasons why we prefer to study the {\sl preliminary} data 
for the asymmetries weighted with a power of $P_{h\perp}$ \cite{HERMES-new} 
instead of the final data for the asymmetries weighted without $P_{h\perp}$ 
\cite{Airapetian:2004tw}, in spite of the caveat that {\sl preliminary} data 
can be subject to changes due to refined data analyses.
Firstly, in the parton model approximation only asymmetries weighted with 
an appropriate power of transverse momentum (e.g., $P_{h\perp}$ in the case 
of Sivers effect in SIDIS) allow a model independent disentanglement of 
transverse parton momenta in the target and in the fragmenting hadron 
\cite{Boer:1997nt}.
An analysis of $A_{UT}^{\sin(\phi-\phi_S)}$ would inevitably be biased
by our prejudice concerning the distribution of transverse parton momenta
in the target and in the fragmentation process, while the use of the
asymmetry (\ref{Eq:03},~\ref{Eq:04}) allows to avoid this problem elegantly.
Secondly, below we will be interested in discussing the Sivers effect 
in DY pair production in the COMPASS experiment at considerably higher 
energies. It has been argued that asymmetries weighted without transverse 
momentum could be subject to strong dilution due to Sudakov effects, 
while this effect could be minimized by weighting the SSA by an appropriate 
power of transverse momentum \cite{Boer:2001he}.

Using the fact that the unpolarized distribution $f_1^a(x)$ and
fragmentation $D_1^a(z)$ functions are well known and parameterized
(see, e.g., \cite{Gluck:1998xa,Martin:2002dr,Kretzer:2001pz}), one could
try to extract directly information on the Sivers function using the
so-called purity method which is being pursued by the HERMES Collaboration
\cite{Tanaka}.
Instead, we choose a different strategy here and fit the HERMES
data \cite{HERMES-new} for which we employ the ansatz
\be\label{Eq:sidis-ansatz}
    xf_{1T\,\SIDIS}^{\perp(1)u}(x) = -xf_{1T\,\SIDIS}^{\perp(1)d}(x)
    = A\,x^B (1-x)^5\;\;,\;\;\;\; xf_{1T\,\SIDIS}^{\perp(1)\bar q}(x)=0 \;.
\ee

Several comments are in order. Firstly, we assume that the Sivers function
for antiquarks can be neglected in comparison to the one for quarks.
Secondly, we imposed the condition (\ref{Eq:th-06}) derived from the
large-$N_c$ limit \cite{Pobylitsa:2003ty}.
Both assumptions are severe constraints, but -- given the size of the
experimental error bars \cite{HERMES-new} -- they can be expected to hold
with sufficient accuracy for our purposes. In fact, we shall see that
the present data are compatible with the ansatz.
Of course, one should expect that future precision data may demand to relax
these constraints. At the present stage, however, these assumptions are very
helpful to reduce the number of unknown quantities to only one, namely the
Sivers $u$-quark distribution. We also neglect strange quark effects.

%
\begin{wrapfigure}{HR}{5.8cm}
\vspace{-0.3cm}
{\epsfxsize=2.4in\epsfbox{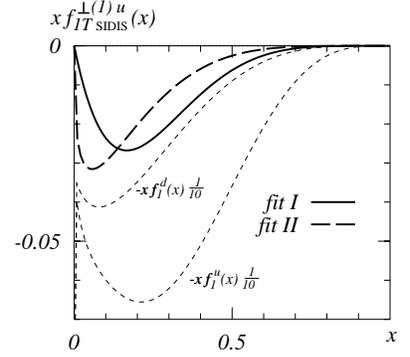}}
\vspace{-0.8cm}
\caption{\footnotesize\sl
    \label{Fig3-Sivers-fit}
    Sivers function according to
    Eqs.~(\ref{Eq:sidis-ansatz},~\ref{Eq:sidis-fit}) as obtained from
    a fit to the HERMES data \cite{HERMES-new},
    see Figs.~\ref{Fig4-Sivers-vs-HERMES}a-c.
    The unpolarized quark distributions $xf_1^q(x)$ at $Q^2=2.5\,{\rm GeV}^2$,
    rescaled by the factor $(-1)/10$, are shown for the sake of comparison.}
\vspace{-0.4cm}
\end{wrapfigure}
\begin{figure}[b!]
\begin{tabular}{ccc}
\hspace{-1cm}   {\epsfxsize=2.4in\epsfbox{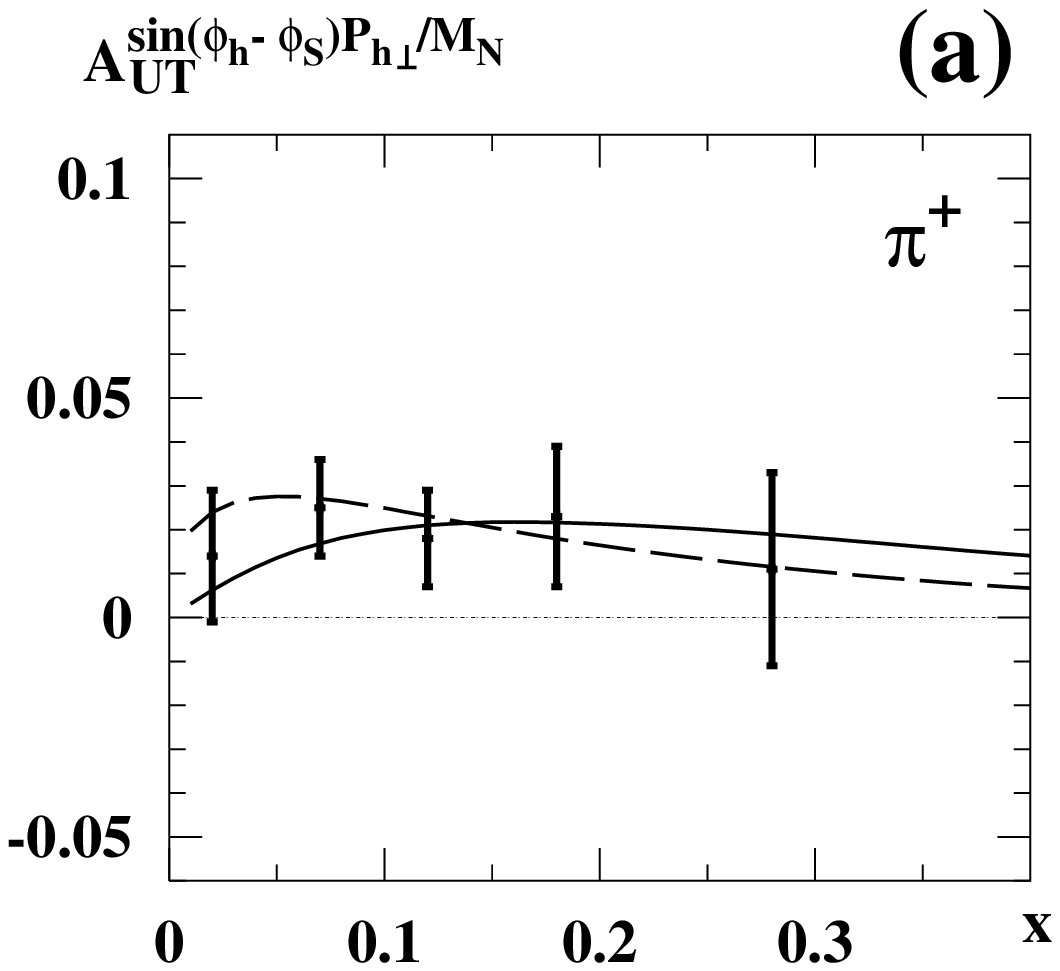}}&
\hspace{-1cm}   {\epsfxsize=2.4in\epsfbox{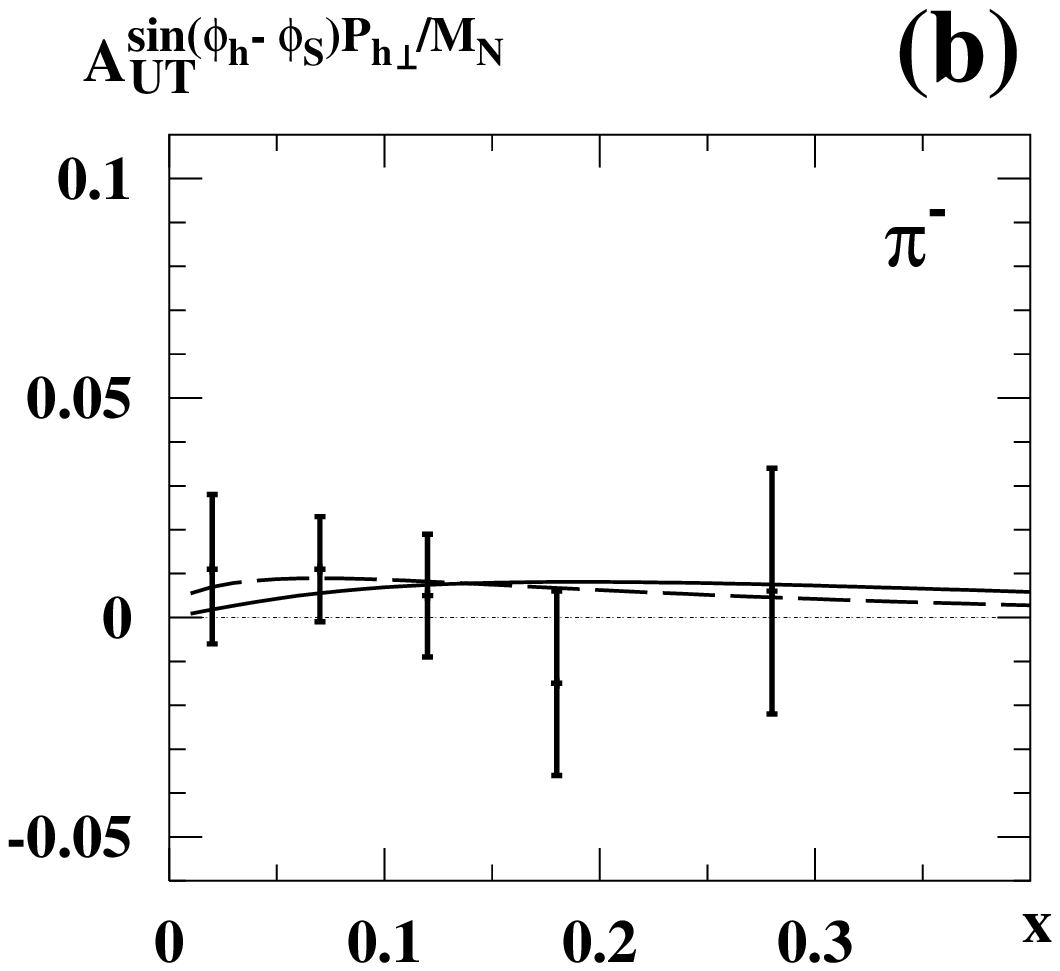}}&
\hspace{-1cm}   {\epsfxsize=2.4in\epsfbox{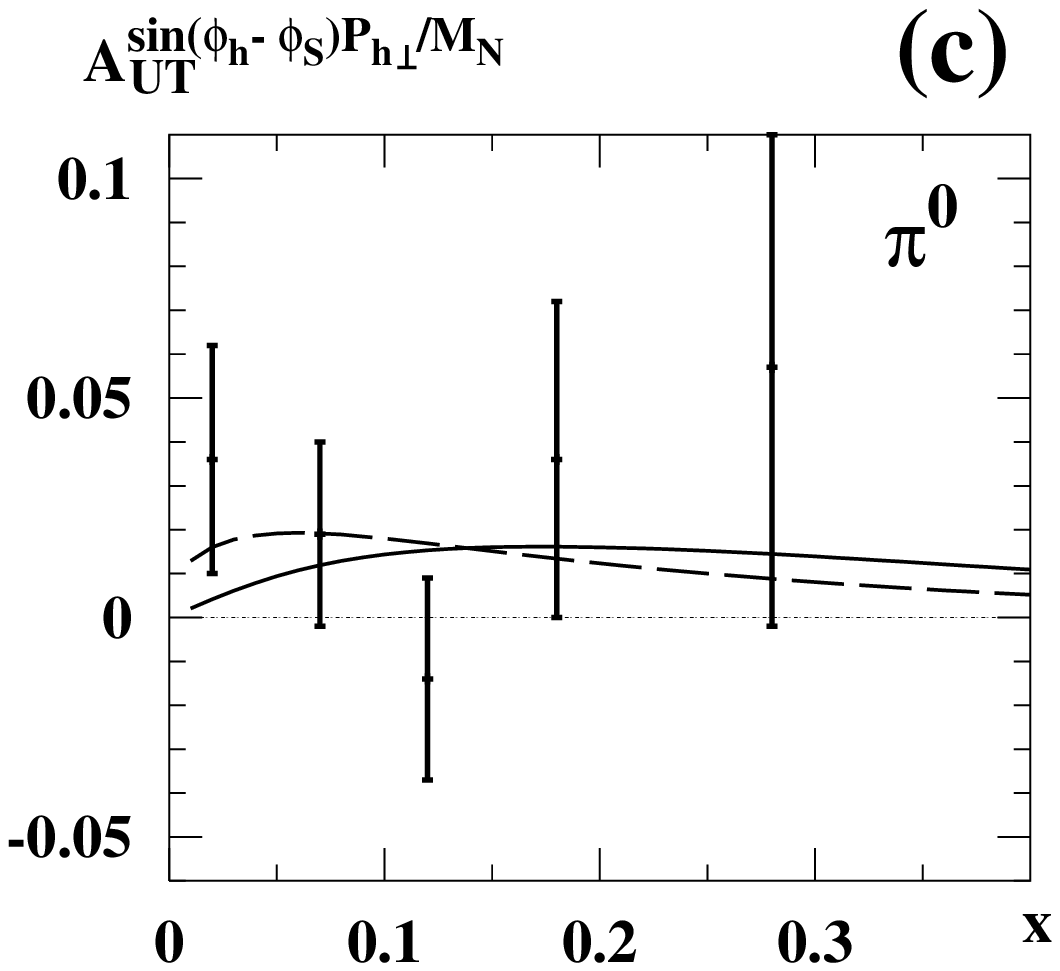}} \\
\hspace{-1cm}   {\epsfxsize=2.4in\epsfbox{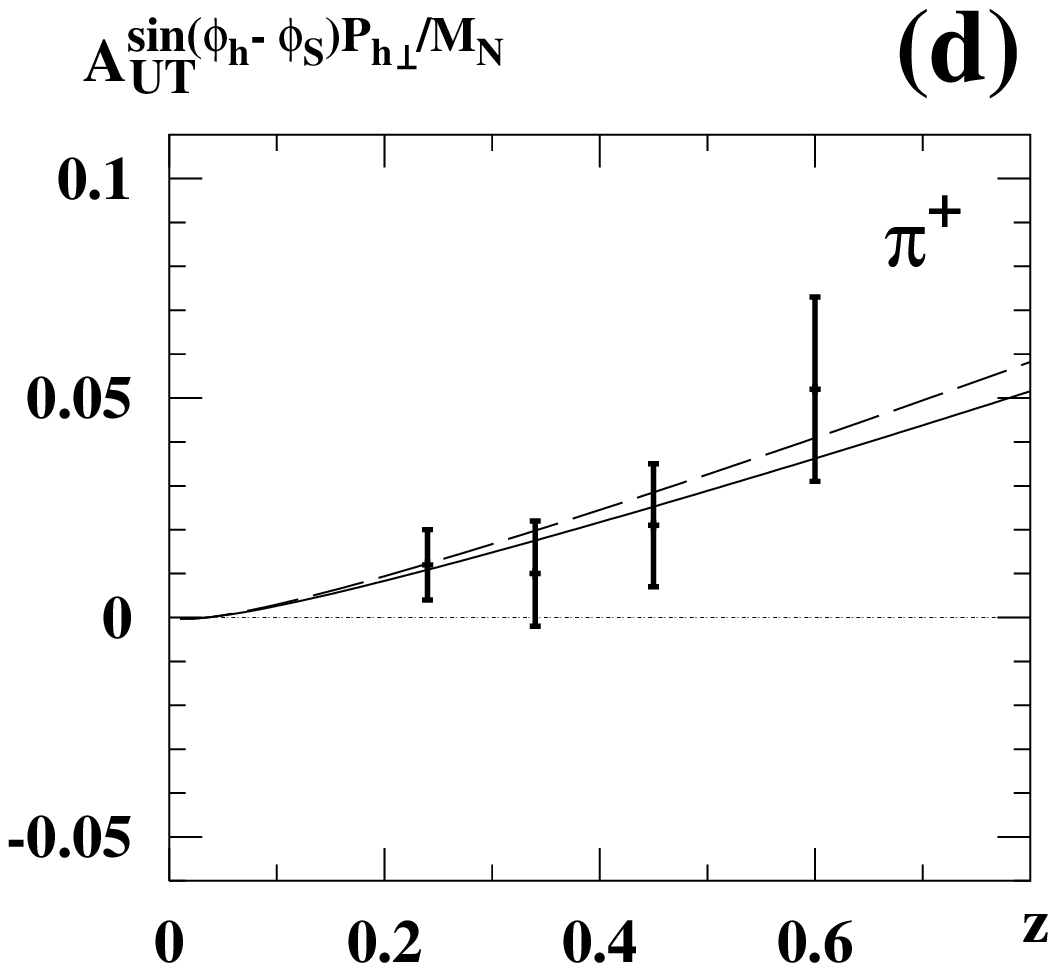}}&
\hspace{-1cm}   {\epsfxsize=2.4in\epsfbox{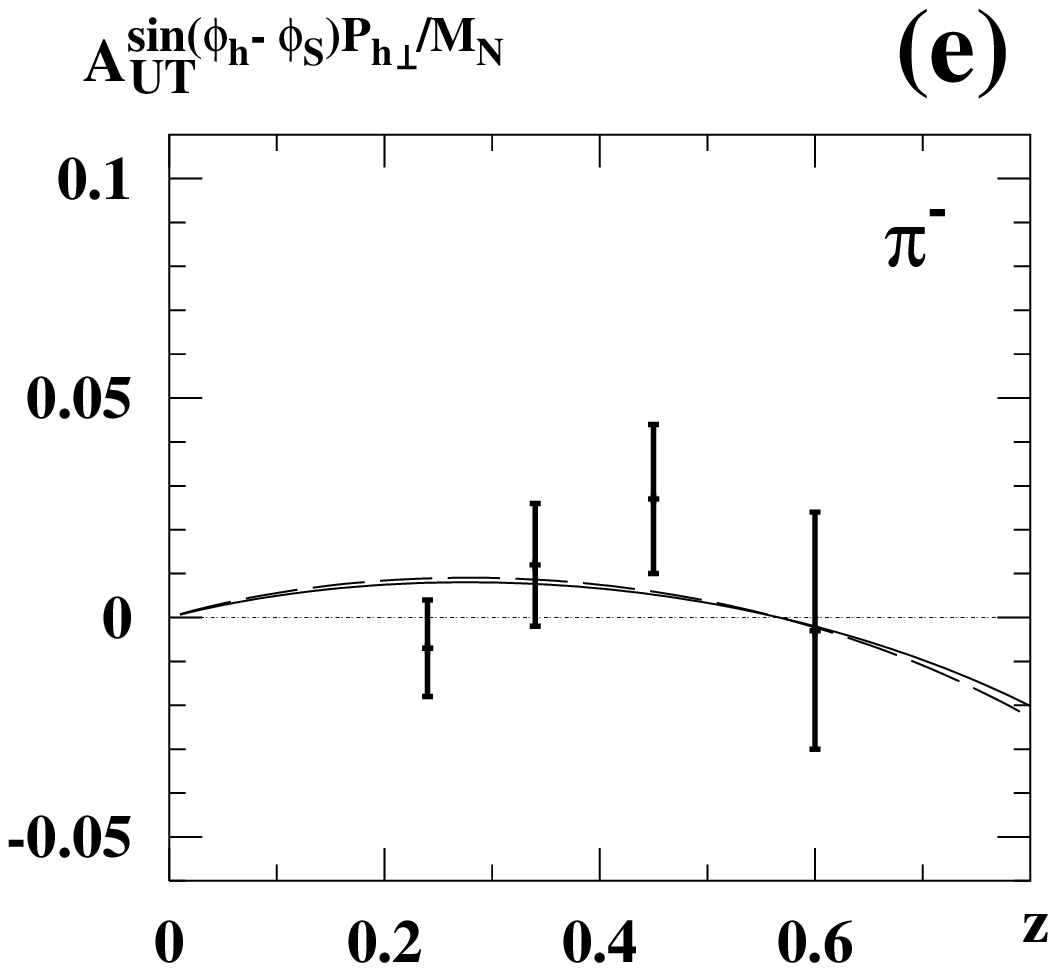}}&
\hspace{-1cm}   {\epsfxsize=2.4in\epsfbox{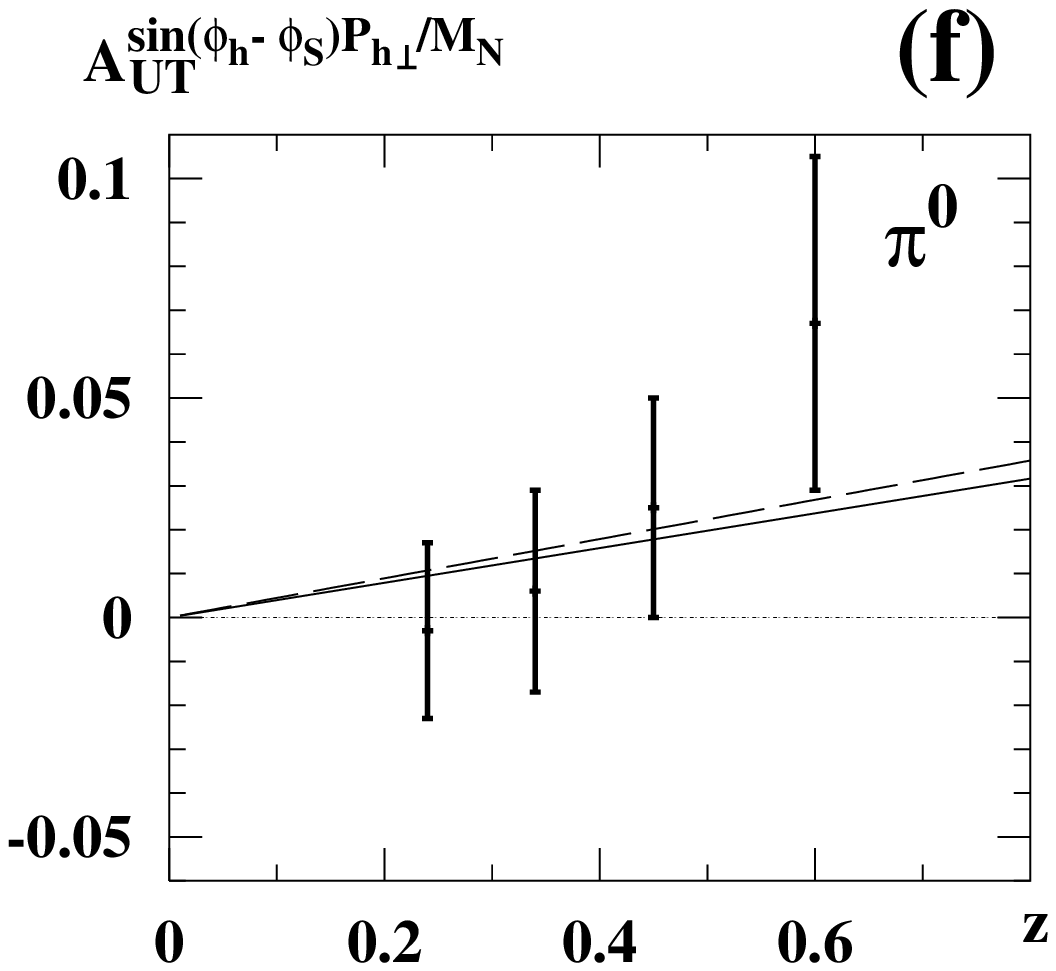}}
\end{tabular}
\vspace{-0.3cm}
\caption{\footnotesize\sl
    \label{Fig4-Sivers-vs-HERMES}
    (a,b,c) 
    The azimuthal SSA $A_{UT}^{\sin(\phi_h-\phi_S)P_{h\perp}/M_N}$ as 
    function of $x$. The preliminary data are from the HERMES experiment 
    \cite{HERMES-new}. The curves are obtained from the large-$N_c$ 
    constrained fits I and II (denoted as in Fig.~\ref{Fig3-Sivers-fit}) 
    of the Sivers function.
    (d,e,f) 
    $A_{UT}^{\sin(\phi_h-\phi_S)P_{h\perp}/M_N}$ as function of $z$, with
    the preliminary data from \cite{HERMES-new}, and the theoretical curves 
    from the fits I and II of the Sivers function. The z-dependent data
    were not used for the fit, and serve as a cross check of our results.}
\end{figure}

In order to illustrate to which extent the HERMES data allow to constrain the
parameters in the ansatz (\ref{Eq:sidis-ansatz}) we performed two fitting
procedures, one
(I) with the parameter $B=1$ fixed from the very beginning, and another one
(II) where both parameters $A$ and $B$ were kept free. 
%
Using for $f_1^a(x)$ and $D_1^a(z)$ the parameterizations \cite{Gluck:1998xa} 
(or \cite{Martin:2002dr} which yields a negligible difference) and
\cite{Kretzer:2001pz} at $Q^2 = 2.5\,{\rm GeV}^2$ we obtain the fits 
\ba
    \mbox{Fit I :}  && xf_{1T\,\SIDIS}^{\perp(1)u}(x)
    = -0.4\,x\,(1-x)^5 \;, \nonumber\\
    \mbox{Fit II :} && xf_{1T\,\SIDIS}^{\perp(1)u}(x)
    = -0.1\,x^{0.3}(1-x)^5 \;, \label{Eq:sidis-fit}
\ea
with a comparable $\chi^2$ per data point of about 0.4.
The fitting functions are of different shape, see Fig.~\ref{Fig3-Sivers-fit},
but they describe the HERMES data \cite{HERMES-new} equally well,
see Figs.~\ref{Fig4-Sivers-vs-HERMES}a-c.
The scale for the Sivers function in Eq.~(\ref{Eq:sidis-fit}) corresponds to
the average scale in the HERMES experiment of $\la Q^2\ra = 2.5\,{\rm GeV}^2$.
We remark that the fits (\ref{Eq:sidis-fit}) are mainly constrained by
the $\pi^+$-data. Leaving the $\pi^0$ (and/or $\pi^-$) data out of the fit
does not affect the numbers in (\ref{Eq:sidis-fit}) significantly.

Thus we see that the experimental accuracy of the data does not allow one to
constrain more sophisticated ans\"atze with more than two free parameters.
Considering the discussion of the large-$x$ behaviour in the previous
section, we have been guided to the ansatz (\ref{Eq:sidis-ansatz}).
However, one should keep in mind that we use this ansatz only in the
region $x<0.4$ covered by HERMES, so the precise shape of 
$f_{1T}^{\perp q}(x)$
in the limit $x\to 1$ is of no relevance for us.

%

Let us confront the results of our fit to the $z$-dependent 
data from \cite{HERMES-new}. Since the latter was not used to constrain 
the fit, the comparison in Figs.~\ref{Fig4-Sivers-vs-HERMES}d-f can be
viewed as a ``cross check'' of the fitting procedure. The expression for
the asymmetry is given by Eq.~(\ref{Eq:04}) but with the average with
respect to $x$ instead of $z$. The shape of the SSA is dictated by the
parameterization for $D_1^a(z)$ from Ref.~\cite{Kretzer:2001pz}.
The asymmetry is linearly rising with $z$ for $\pi^0$ 
(where $D_1^q$ is the same for all $q=u,\,\bar u,\, d,\,\bar d$) and 
nearly so for $\pi^+$ (where favoured flavour approximation works well), 
but it has a peculiar shape for $\pi^-$ (where $1/N_c$-corrections 
to the Sivers function would have the most impact). 
We conclude that the ansatz (\ref{Eq:sidis-ansatz}) and the fits 
(\ref{Eq:sidis-fit}) are well compatible with the z-dependence of 
the data, see Figs.~\ref{Fig4-Sivers-vs-HERMES}d-f.

We observe that the obtained fit satisfies
$|f_{1T\,\SIDIS}^{\perp\,(1)\,a}(x)|<\frac{1}{10}\,f_1^a(x)$,
see Fig.~\ref{Fig3-Sivers-fit}. Multiplying Eq.~(\ref{Eq:th-02}) by
$|{\bf p}_T|/(2M_N)$
and integrating it over transverse momenta gives the inequality
$|f_{1T\,\SIDIS}^{\perp(1)a}(x)|\le\frac{\la p_T\ra}{2M_N}\,f_1^q(x)$
which (defines a phenomenological mean parton transverse momentum
$\la p_T\ra$ and) is less restrictive than the inequality observed in
Fig.~\ref{Fig3-Sivers-fit}, if we assume $\la p_T\ra\approx0.8\,{\rm GeV}$
(see Ref.~\cite{D'Alesio:2004up}). In this sense, we note that our result is
in agreement with the positivity bound in Eq.~(\ref{Eq:th-02}).

Let us also remark that the HERMES data \cite{HERMES-new} are compatible with
the large-$N_c$ counting rule in Eq.~(\ref{Eq:th-06}) within their present
statistical accuracy, which is proven by the fact that a fit with the ansatz
(\ref{Eq:sidis-ansatz}) works. The sum rule (\ref{Eq:th-05}) is satisfied
by our parameterization  (recall the suppression of the gluon
Sivers function in the large-$N_c$ limit).

Experiments with the deuterium target, for which\footnote{ 
	We neglect nuclear binding effects and assume isospin symmetry
 	which is legitimate given the present level of accuracy.
	Parton distributions without a target label 
	$(D=\mbox{deuteron},\;p=\mbox{proton},\;n=\mbox{neutron})$ refer, 
	as everywhere in this note, to the proton.}
$f_{1T}^{\perp u/D} \approx f_{1T}^{\perp u/p}+f_{1T}^{\perp u/n} =$
$f_{1T}^{\perp u} + f_{1T}^{\perp d}$, etc., are best suited to study 
deviations from the ansatz (\ref{Eq:sidis-ansatz}).
Thus, the Sivers effect spin asymmetries from the deuterium target 
are suppressed with respect to proton asymmetries by a power of $N_c$ 
in the large-$N_c$ limit. The preliminary COMPASS data on 
$A_{UT}^{\sin(\phi-\phi_S)}$ from a deuterium target are compatible with 
zero within error bars \cite{Bradamante:2004qh}, and thus do not contradict 
the large-$N_c$ motivated ansatz (\ref{Eq:sidis-ansatz}).

Let us compare our result to parameterizations of the Sivers function in
the literature obtained from studies of SSA in $p^\uparrow p\to\pi X$. The
fits in Eq.~(\ref{Eq:sidis-fit}) agree in sign, but are 
shifted towards smaller $x$, and one order of magnitude larger than the 
Sivers function obtained in Ref.~\cite{Anselmino:1998yz}.
On the basis of the latter it was estimated \cite{Efremov:2003tf} that the 
Sivers effect can be neglected with respect to the Collins effect in the 
twist-3 SSA $A_{UL}^{\sin\phi}$ observed at HERMES \cite{Airapetian:1999tv}.
Our considerably more sizeable result does not support these conclusions
and indicates a possible necessity to reconsider the
interpretation \cite{Efremov:2001cz,DeSanctis:2000fh,Ma:2002ns} of the
HERMES data on $A_{UL}^{\sin\phi}$, though recent studies indicate that the
Sivers effect cannot play a dominant role in this SSA \cite{Anselmino:2004ht}.

In the earlier extractions \cite{Anselmino:1994tv,Anselmino:1998yz}
$f_{1T}^{\perp(1)q}(x)$ was so small due to a small value of
$\la p_T\ra < 0.2\,{\rm GeV}$ assumed in these references.
With the updated value $\la p_T\ra\approx0.8\,{\rm GeV}$ the resulting
updated Sivers functions \cite{D'Alesio:2004up} which fits the FNAL data
\cite{Adams:1991rw} is of comparable magnitude as our result.
However, the 
different $x$-shape 
of the $f_{1T\,\SIDIS}^{\perp(1)q}$ required to describe HERMES data as 
compared to the $f_{1T}^{\perp(1)q}$ required to describe the FNAL data 
remains.

Finally we note that the prediction of the sign of the Sivers function
for $u$- and $d$-flavour in Eq.~(\ref{Eq:th-07}) is confirmed by our
parameterization, which suggests that the physical picture of the Sivers
effect of Ref.~\cite{Burkardt:2002ks} is apparently able to catch main
features of the Sivers effect.

\section{Sivers effect in the Drell-Yan process}
\label{Sec-4:Sivers-in-DY}

The information on $f_{1T}^\perp$ deduced in Sec.~\ref{Sec-3:Sivers-in-SIDIS}
from the HERMES data \cite{HERMES-new} is rough, however, as we shall see,
sufficient for our goal to predict the sign and to gain insight into the
magnitude of the SSA in DY.

The process $p^\uparrow h\to \mu^+\mu^-X$ (with $h=\bar p,\,\pi^-$ in the
following) is characterized by the variables $s=(p_1+p_2)^2$,
the dilepton invariant mass $Q^2=(k_1+k_2)^2$ with $p_{1/2}$ (and $k_{1/2}$)
indicating the momenta of the incoming proton and hadron $h$
(and the outgoing lepton pair), and the rapidity
\be
    y=\frac12\,{\rm ln}\frac{p_1(k_1+k_2)}{p_2(k_1+k_2)} \;.
\ee
Let us consider the azimuthal SSA which is weighted by $|{\bf q}_T|$,
the dilepton momentum transverse with respect to the collision axis, and
defined as a sum over the events $i$ according to
\be
    A_{UT}^{\sin(\phi-\phi_S)\frac{q_T}{M_N}} =
    \frac{\sum_i\sin(\phi_i-\phi_{S,i})\frac{|{\bf q}_{T,i}|}{M_N}
    \{N^\uparrow(\phi_i,\phi_{S,i})-N^\downarrow(\phi_i,\phi_{S,i}+\pi)\}}
    {\frac12\sum_i
    \{N^\uparrow(\phi_i,\phi_{S,i})+N^\downarrow(\phi_i,\phi_{S,i}+\pi)\}} \;,
\ee
where $\uparrow(\downarrow)$ denote the transverse polarization of the
proton. (See Fig.~\ref{fig2-processes-kinematics}b for the definition of
the kinematics.) To leading order the SSA is given by
\be\label{Eq:ATT-0}
    A_{UT}^{\sin(\phi-\phi_S)\frac{q_T}{M_N}}(y,Q^2) = 2\;\frac{
    \sum_a e_a^2 \, x_1f_{1T\,\DY}^{\perp(1) a/p}(x_1)\,x_2f_1^{\bar a/h}(x_2)}{
    \sum_a e_a^2 \, x_1f_1^{a/p}(x_1)\,                 x_2f_1^{\bar a/h}(x_2)} \;,
\ee
where the parton momenta $x_{1/2}$ in Eq.~(\ref{Eq:ATT-0})
are fixed in terms of $s$, $Q^2$ and $y$,
\be\label{Eq:x12}
    x_{1/2} = \sqrt{\frac{Q^2}{s}}\,e^{\pm y}\;.\ee
The sums in Eq.~(\ref{Eq:ATT-0}) run over all quark and antiquark flavours,
and we indicate explicitly to which hadron the distributions refer.

\begin{figure}[t]
\begin{tabular}{cc}
    \hspace{1.5cm}
    {\epsfxsize=2.4in\epsfbox{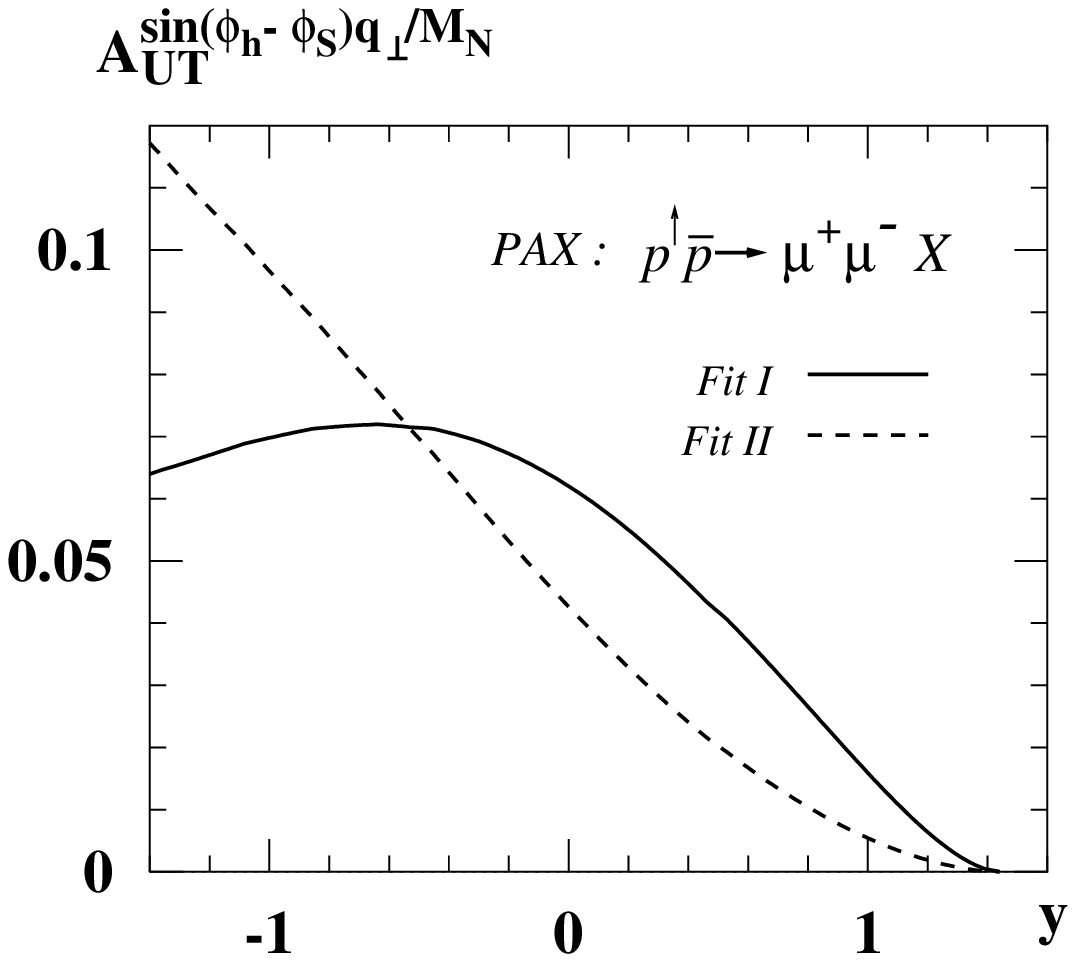}}
    {\epsfxsize=2.4in\epsfbox{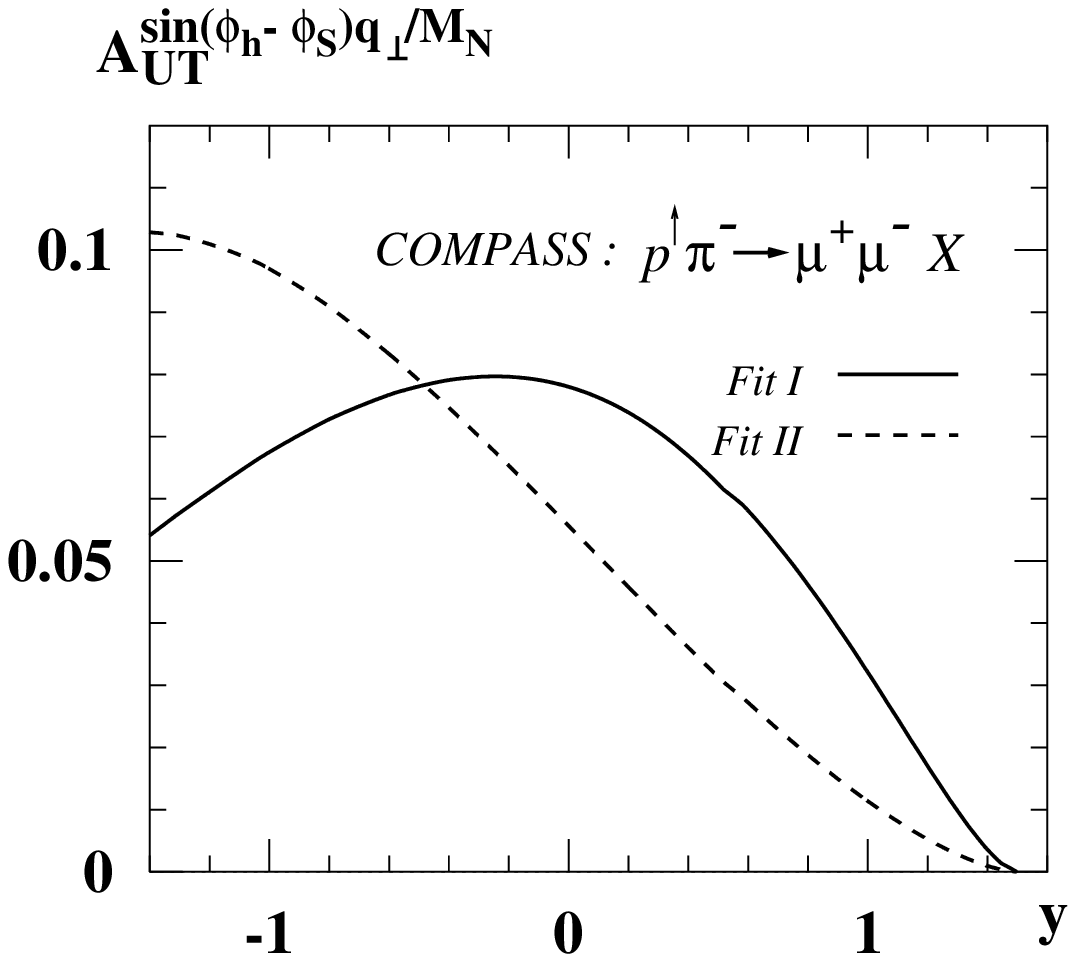}}
\end{tabular}
\vspace{-0.3cm}
\caption{\footnotesize\sl
    \label{Fig5-DY}
    The azimuthal SSA $A_{UT}^{\sin(\phi_h-\phi_S)q_\perp/M_N}$ in
    Drell-Yan lepton pair production, $p^\uparrow h\to\mu^+\mu^- X$,
    as function of $y$: (a) for the kinematics of the PAX experiment
    where the hadron $h=\bar{p}$, (b) for the kinematics of the
    COMPASS experiment where $h=\pi^-$. The different curves correspond
    to the fits I and II (see Eq.~(\ref{Eq:sidis-fit})), including the
        sign-reversal in~(\ref{Eq:01}).}
\end{figure}

In the PAX experiment antiprotons with a beam energy of $25\,{\rm GeV}$
could be available, i.e.\   $s=45\,{\rm GeV}^2$.
In this kinematics one could explore the region around $Q^2=2.5\,{\rm GeV}^2$,
which is below the region of $J/\psi$ production, and well above
the region of dileptons from $\Phi(1020)$-decays.
Taking into account the change of sign in the Sivers function in DY
as compared to SIDIS, see Eq.~(\ref{Eq:01}), we obtain the result shown in
Fig.~\ref{Fig5-DY}a.\footnote{The DY asymmetry appears positive
            like the SIDIS asymmetry at HERMES, despite the change of
            sign of the Sivers function due to conventions: In DY we define
            the z-axis in the direction in which the polarized particle moves.
            In SIDIS at HERMES the z-axis is defined in the opposite direction,
            see Figs. 2a and 2b.}
We observe that the two fits I and II, which describe the HERMES data
of SIDIS equally well, give clearly distinguishable results in DY.
Considering depolarization, detector acceptance and other effects,
it might be difficult to distinguish the effect of the different
parameterizations in Eq.~(\ref{Eq:sidis-fit}). However, the asymmetry is
large enough to check unambiguously the QCD prediction of the different
sign of the Sivers function in DY and SIDIS.

%
%

In the COMPASS experiment using a $\pi^-$ beam ($s=400\,{\rm GeV}^2$)
one could also measure the asymmetry (\ref{Eq:ATT-0}).
In Fig.~\ref{Fig5-DY}b we show the asymmetry for $Q^2=20\,{\rm GeV}^2$
using for the pion the parameterization from Ref.~\cite{Gluck:1999xe}.
Although $f_1^{a/\pi}(x)$ is far less constrained by data compared to
$f_1^{a/p}(x)$ the result in Fig.~\ref{Fig5-DY}b is rather insensitive to
the choice of parameterization, and changes very little if we use the pion
distributions of Ref.~\cite{Sutton:1991ay} (consistently in combination with
the nucleon distributions from Ref.~\cite{Martin:2002dr}).
We observe a situation, which is qualitatively and quantitatively similar
to the case of DY from $p\bar p$-collisions.
Note that we neglected evolution effects (from $Q^2_0=2.5\,{\rm GeV}^2$
in Eq.~(\ref{Eq:sidis-fit}) to $Q^2=20\,{\rm GeV}^2$ in Fig.~\ref{Fig5-DY})
for the Sivers function.
However, the influence of evolution is presumably much smaller
than other uncertainties in our study.
Note that by using the $q_\perp$-weighted SSA we have avoided
another serious problem in this context, namely Sudakov suppression 
\cite{Boer:2001he}, see the remarks in the previous section.

In order to extract quantitative information from the future COMPASS and PAX
experiments it is necessary to go beyond the LO formalism, to consider effects
of soft gluons and K-factors, and to study the role of possible higher twist
effects. The corrections due to these effects
cannot be expected to be negligible. However, they are unlikely to be
able to change the sign of the asymmetry. Thus, both the COMPASS as well as
the PAX experiment could provide a thorough experimental test of the QCD
prediction in Eq.~(\ref{Eq:01}).

SSA in DY can also be studied at RHIC in $p^\uparrow p\to \mu^+\mu^- X$.
Since only one proton needs to be polarized the counting rates would be somehow
more sizeable than in the case of double spin asymmetries related to the
transversity distribution $h_1^a(x)$ which are, however, small
\cite{Bunce:2000uv}. Moreover, in this case, one is 
sensitive to the Sivers antiquark distribution which is not constrained by
the HERMES data. 
We remark that the RHIC experiment is well suited to learn, e.g., about the 
Sivers function from SSA in $p^\uparrow p\to \pi X$ \cite{Adams:2003fx} or 
the gluon Sivers function \cite{Boer:2003tx,Anselmino:2004nk}.

\section{Conclusions}
\label{Sec-5:summary}

The recently reported HERMES data \cite{HERMES-new,Airapetian:2004tw} on
SSA provide a theoretically unambiguous experimental evidence for
the existence of T-odd distribution (and fragmentation) functions.
We analyzed the HERMES data and demonstrated that they are consistent with
predictions from the large-$N_c$ limit of QCD \cite{Pobylitsa:2003ty} for
the Sivers functions, namely $f_{1T}^{\perp u} = - f_{1T}^{\perp d}$ modulo
$1/N_c$ corrections. Imposing this large-$N_c$ result as an exact constraint
we were able to obtain parameterizations of the Sivers quark distribution
functions. The neglect of $1/N_c$ corrections (as well as antiquark effects)
in a first approximation is reasonable, keeping in mind the large error
bars of the present data which do not allow to constrain more sophisticated
ans\"atze.

On the basis of the obtained parameterizations we estimated SSA in the
Drell-Yan process for the PAX $(p^\uparrow \bar p\to \mu^+\mu^-X)$ and
COMPASS $(p^\uparrow \pi^-\to \mu^+\mu^-X)$ experiment.
According to the theoretical understanding of T-odd parton distributions
in QCD the Sivers function should obey a particular universality relation,
namely appear with opposite sign in DY and SIDIS \cite{Collins:2002kn}. Our
estimates show that both experiments could be able to test this prediction,
which would be a crucial check of the present understanding of T-odd
distribution functions and the QCD factorization approach to the description
of SSA.

\paragraph{Acknowledgement.}
The work is partially supported by BMBF and DFG of Germany, the
COSY-Juelich project, the Transregio Bonn-Bochum-Giessen, and is
part of the European Integrated Infrastructure Initiative Hadron
Physics project under contract number RII3-CT-2004-506078. A.E.
is supported by grants RFBR 03-02-16816 and DFG-RFBR 03-02-04022.

\paragraph{Note added:}
After this manuscript was completed the COMPASS collaboration has
published final data on transverse target SSA from a deuteron target
\cite{Alexakhin:2005iw}.  We stress that the large-$N_c$ prediction for
the flavour dependence of the Sivers function \cite{Pobylitsa:2003ty}
naturally explains the
compatibility of the sizeable Sivers SSA from a {\sl proton} target
observed at HERMES \cite{HERMES-new,Airapetian:2004tw} and the small
(consistent with zero within error bars) SSA from a {\sl deuteron} target
observed at COMPASS \cite{Bradamante:2004qh,Alexakhin:2005iw}. The Sivers
effect in the deuteron is sensitive to the flavour combination
$(f_{1T}^{\perp u}+f_{1T}^{\perp d})$, and thus suppressed with respect to
the Sivers effect in the proton by one order of $N_c$ in the large-$N_c$
expansion. In nature $N_c=3$ is sufficiently large to explain the
observations -- considering the statistics of the first experiments
\cite{HERMES-new,Airapetian:2004tw,Bradamante:2004qh,Alexakhin:2005iw}.

\end{document}